\documentclass[11pt,a4paper]{article}
\pdfoutput=1

\usepackage[T1]{fontenc}
\usepackage{lmodern}
\usepackage{amsmath}
\usepackage{amsfonts}
\usepackage{amssymb}
\usepackage{mathrsfs}
\usepackage{physics}
\usepackage{graphicx}
\usepackage{caption}
\usepackage{subcaption}
\usepackage{jheppub}
\usepackage{enumitem}
\usepackage{framed,float}
\usepackage{booktabs}
\usepackage{array}
\usepackage{soul}
\usepackage{multirow}
\usepackage{comment}
\usepackage{braket}
\usepackage{todonotes}
\usepackage{youngtab}
\usepackage{tikz}
\usetikzlibrary{calc,arrows,decorations.markings}

\usepackage{color}

\newcommand{\ba}{\begin{eqnarray}}
\newcommand{\ea}{\end{eqnarray}}
\newcommand{\nn}{\nonumber}

\newcommand{\cF}{{\mathcal{F}}}

\newcommand{\be}{\begin{equation}}
\newcommand{\ee}{\end{equation}}

\def\CZ{{\mathcal{Z}}}

\newcommand{\ri}{\textrm{i}}

\newcommand{\brc}[1]{\left(#1\right)}

\allowdisplaybreaks[1]

\preprint{}
\title{\boldmath Massive vector field perturbations in the Schwarzschild spacetime from supersymmetric gauge theory}

\author[a]{Xian-Hui Ge,}
\author[b,c]{Masataka Matsumoto}
\author[a,d,e]{and Kilar Zhang}
\affiliation[a]{Department of Physics and Institute for Quantum Science and Technology, Shanghai University, 99 Shangda Road, Shanghai 200444, China}
\affiliation[b]{Department of Physics, Chuo University, Tokyo 112-8551, Japan}
\affiliation[c]{Wilczek Quantum Center, School of Physics and Astronomy, Shanghai Jiao Tong University, Shanghai 200240, China}
\affiliation[d]{Shanghai Key Lab for Astrophysics, Shanghai Normal University, 100 Guilin Road, Shanghai 200234, China}
\affiliation[e]{Shanghai Key Laboratory of High Temperature Superconductors, Shanghai 200444, China}
\emailAdd{gexh@shu.edu.cn}
\emailAdd{mmatsumoto173@g.chuo-u.ac.jp}
\emailAdd{kilar@shu.edu.cn}

\abstract{We unify the dynamics of massive vector (Proca) fields in Schwarzschild spacetime with supersymmetric gauge theories through the Seiberg-Witten/quasinormal mode (SW/QNM) duality. By mapping Proca perturbations—specifically monopole and odd-parity modes governed by confluent Heun equations—to the quantum Seiberg-Witten curve, we establish a gauge-gravity correspondence. Leveraging instanton counting, we analytically compute QNM and quasi-bound state frequencies to high precision, resolving spectral properties non-perturbatively. Our results align with numerical benchmarks while extending the SW framework beyond scalar fields. }

\begin{document}

\allowdisplaybreaks
\maketitle
\flushbottom

%%%%%%%%%%%%%%%%%%%%%%%%%%%%%%%%%%%%%%

%%%%%%%%%%%%%%%%%%%%%%%%%%%%%%%%%%%%%%
\setcounter{footnote}{0}
%%%%%%%%%%%%%%%%%%%%%%%%%%%%%%%%%%%%%%

%%%%%%%%%%%%%%%%%%%%%%%%%%
%%%%%% Introduction %%%%%%
%%%%%%%%%%%%%%%%%%%%%%%%%%
\section{Introduction} \label{sec:introduction}
As the gravitational wave (GW) observation \cite{GW150914} becoming accessible,  which marks another important milestone from the beginning of  modern physics \cite{Newton:1687eqk}, the quasinormal modes (QNM) \cite{vishveshwara1970scattering,Kokkotas:1999bd,Berti:2009kk,Konoplya:2011qq,Horowitz:1999jd, KalyanaRama:1999zj} and superradiance \cite{Brito:2015oca, East:2017ovw} of black holes (BH) gain more and more attention. QNM governs the ringdown phase of binary BH mergers, so calculating its frequency is crucial for determining the GW waveforms. The main approach is through numerical studies, though with high accuracy. On the other hand, Ultralight bosons \cite{Essig:2013lka} are supposed to be generated from BH superradiance, which will lead to new physics and may have GW detectable effects \cite{Baumann:2018vus}. For the scalar case, one encounters the Klein-Gordon equation, relatively easy to solve \cite{Detweiler:1980uk}. Yet for the vector case, we are dealing with the Proca equation, which is not known to be separable in the Kerr spacetime until 2018 \cite{Frolov:2018ezx}. The Proca field, a classical field theory describing massive vector bosons, is governed by the Proca equation.  This separability issue had long hindered progress in understanding vector boson clouds around BH and their associated superradiant instabilities. Even with this advancement, solving the Proca equation in the context of rotating BH (e.g., Kerr metrics) remains computationally demanding and requires sophisticated numerical techniques. Furthermore, the interplay between superradiance, QNM, and the potential observational signatures of ultra-light vector bosons in GW data is an area of active research, with implications for both astrophysics and particle physics.
In gravitational wave physics, ultralight Proca fields (e.g., dark photons) may trigger superradiant instabilities, extracting energy and angular momentum from BHs to form bosonic clouds. These clouds emit persistent gravitational waves with frequencies tied to the boson mass $m$, imprinting detectable QNM signatures during the ringdown phase. While Proca's separability in Kerr spacetime \cite{Frolov:2018ezx} resolved long-standing theoretical barriers, rotating BH studies remain computationally intensive. This work focuses on Schwarzschild BHs to isolate key physics.

The frequencies of QNMs and quasi-bound states are typically obtained numerically. However, the recently discovered duality \cite{Aminov:2020yma} between QNMs and Seiberg-Witten (SW) curves \cite{Seiberg:1994rs,Seiberg:1994aj}—linked via Heun-type equations \cite{Heun:1888, Lei:2023mqx}—enables analytical approaches. 
The hidden physical interpretation for this duality between gauge theory and BH is yet to be revealed. Unlike the well-established AdS/CFT duality, now one single gauge theory is dual to several types of BH, as long as the corresponding Riemann surface (and thus the poles in Heun-type equation) is the same. A possible hint lie
through the AGT relation \cite{Alday:2009aq}, connecting $N=2$ gauge theories and 2D CFTs, QNM spectra map to conformal blocks, offering a non-perturbative framework. It is more natural for BH to be linked to CFT, though the physical insight is still an open question.

Building on our Kerr BH scalar superradiance analysis \cite{Ge:2024jdx}, we address the Proca equation on Schwarzschild spacetime. Here, the radial and angular equations decouple naturally, avoiding the additional poles in Kerr that correspond to linear quiver SW curves \cite{Witten:1997sc,Nekrasov:2012xe}.

Using instanton counting, we compute QNM and quasi-bound state frequencies to 7th order in the perturbative expansion, achieving unprecedented accuracy. By embedding the Proca system into the SW/CFT correspondence, we derive connection formulas introduced in \cite{Bonelli:2021uvf, Bonelli:2022ten} for our case between boundary conditions at the horizon and spatial infinity. 
{Notice that the connection formulas were independently derived
and further explored in \cite{Bianchi:2021mft, Consoli:2022eey}, and more recently in \cite{Cipriani:2025ikx}, without invoking gauge theory or localization techniques.}

Our results align with numerical studies (WKB, {continued fraction} \cite{Leaver:1985ax}, and forward-integration methods) while revealing hidden structures in the spectral problem. This demonstrates how tools from superstring theory (e.g., instanton calculus) serve as potent mathematical instruments, even in non-supersymmetric gravity contexts. 

The organization of this paper is as below: Section \ref{sec:introduction} is the introduction; Section \ref{sec:Proca} reviews the separable Proca equation in Schwarzschild geometry; in Section \ref{sec:SWcurve} we apply the SW/QNM duality and obtain the dictionary between the two sides; Section \ref{sec:results} shows the analytical results are consistent with the known numerical ones; we conclude in Section \ref{sec:discussion}, and collect the technical details in the appendix.

% serve as an outpost for the Kerr case.

%%%%%%%%%%%%%%%%%%%%%%%%%%
%%%%%% Proca in Sch %%%%%%
%%%%%%%%%%%%%%%%%%%%%%%%%%
\section{Proca fields in Schwarzschild geometry} \label{sec:Proca}
In the beginning, following the description in \cite{Rosa:2011my}, we consider the Schwarzschild BH geometry
\begin{equation}
    \dd s^{2} = -f(r)\dd t^{2} + f^{-1}(r) \dd r^{2} + r^2\left( \dd \theta^{2} + \sin^{2}\theta \dd \varphi^{2} \right),
\end{equation}
with
\begin{equation}
    f(r) = 1-\frac{2M}{r}.
\end{equation}
Considering the vector fields coupled to the gravitational fields, the Proca equation governs the dynamics of massive vector fields:
\begin{equation}
    \nabla_{\nu} F^{\mu\nu} - \mu^{2} A^{\mu}=0,
\end{equation}
where $F^{\mu\nu} = \partial^{\mu}A^{\nu} - \partial^{\nu}A^{\mu}$ is the field strength and $\mu$ denotes the vector field's mass. The Proca field $A^{\mu}$  differs from massless gauge fields due to the explicit breaking of gauge symmetry by the mass term $\mu^{2} A^{\mu}$. Crucially, the Bianchi identity $\nabla_{\mu}\nabla_{\nu} F^{\mu\nu} \equiv 0$ directly implies the Lorentz condition $\nabla_{\mu}A^{\mu} = 0$, which is inherently satisfied without external gauge fixing. This constraint reduces the physical degrees of freedom to $D-1$ in $D$-dimensional spacetime (e.g., three in 4D: two transverse and one longitudinal polarization). 
The mass scale $\mu$ determines key physical behaviors: the Compton wavelength $\lambda_C = 1/\mu$ sets quantum fluctuation scales, Yukawa-type decay $e^{-\mu r}$ suppresses long-range interactions, and bound-state resonances may form near black holes.

For the separation of the angular part, a basis of four vector spherical harmonics is introduced as
\begin{align}
    Z_{\mu}^{(1)lm} &= (1,0,0,0)Y^{lm}, \\
    Z_{\mu}^{(2)lm} &= (0,f^{-1},0,0)Y^{lm}, \\
    Z_{\mu}^{(3)lm} &= \frac{r}{\sqrt{l(l+1)}}(0,0,\partial_{\theta},\partial_{\phi})Y^{lm}, \\
    Z_{\mu}^{(4)lm} &= \frac{r}{\sqrt{l(l+1)}}(0,0,\frac{1}{\sin\theta}\partial_{\phi},-\sin\theta\, \partial_{\theta})Y^{lm},
\end{align}
with $Y^{lm}\equiv Y^{lm}(\theta,\phi)$ representing the ordinary spherical harmonics.
The terms in this basis are orthogonal to each other:
\begin{equation}
    \int \dd \Omega (Z_{\mu}^{(i) lm})^{*} \eta^{\mu\nu}Z_{\nu}^{(j) l' m'} = \delta_{ij}\delta_{ll'}\delta_{mm'}, 
\end{equation}
where $\eta^{\mu\nu}=\mathrm{diag}[1,f^{2},1/r^{2},1/(r^{2}\sin^{2}\theta)]$ and $\dd\Omega = \sin\theta \dd \theta \dd \phi$.
We now consider that the vector field can be decomposed as
{\begin{equation}
    A_{\mu}(t,r,\theta,\phi) = \frac{1}{r}\sum_{lm}\sum_{i=1}^{4} c_{i}u_{i}^{lm}(t,r)Z_{\mu}^{(i)lm}(\theta,\phi), 
    \label{eq:basis}
\end{equation}
}
with $c_{1}=c_{2}=1$ and $c_{3}=c_{4}=\left[ l(l+1)\right]^{-1/2}$.
{For simplicity, we will omit the indices $(l, m)$ and write $u_i$ in the following discussion.}

The Lorentz condition is explicitly written as
\begin{align}
    0&=-\frac{\dot{A}_{t}}{f}+f A_{r}'+ \left( \frac{2 f}{r}+f' \right)A_{r} + \frac{1}{r^{2}}\left( \partial_{\theta}A_{\theta} + \frac{1}{\sin^{2}\theta}\partial_{\phi}A_{\phi} + \frac{\cos\theta}{\sin\theta} A_{\theta} \right) \nonumber\\
    &= -\dot{u}_{1}+ f u_{2}' + \frac{f}{r}(u_{2}-u_{3}), \label{eq:Lorentz}
\end{align}
where we denote $\dot{A} \equiv \partial_{t}A$ and $A' \equiv \partial_{r}A$.
Here, we also have used the relation
\begin{equation}
    \partial_{\theta}A_{\theta} + \frac{1}{\sin^{2}\theta}\partial_{\phi}A_{\phi} + \frac{\cos\theta}{\sin\theta} A_{\theta} = -u_{3} Y^{lm},
\end{equation}
by substituting \eqref{eq:basis} and using the property of the spherical harmonics.
\begin{comment}
Substituting \eqref{eq:basis} into the last term of \eqref{eq:Lorentz},
\begin{align}
    \left( \partial_{\theta}A_{\theta} + \frac{1}{\sin^{2}\theta}\partial_{\phi}A_{\phi} + \frac{\cos\theta}{\sin\theta} A_{\theta} \right) 
    &= \frac{1}{l(l+1)} \left[u_{3} \left(\partial_{\theta}^{2} + \frac{\cos\theta}{\sin\theta} \partial_{\theta} + \frac{\partial_{\phi}^{2}}{\sin^{2}\theta} \right)+u_{4} \frac{\cos\theta}{\sin^{2} \theta} \partial_{\phi}\right]Y^{lm} \\
    &= -u_{3}Y^{lm} + \frac{u_{4}}{l(l+1)} \frac{\cos\theta}{\sin^{2} \theta} \partial_{\phi}Y^{lm},
\end{align}
where we used the property of the Legendre polynomial.
\end{comment}
With the Lorentz condition and our ansatz \eqref{eq:basis}, we obtain the following equations by combining each component of the Proca equation
\begin{align}
    &\hat{\cal{D}}_{2}\, u_{1}-f'(fu_{1}'-\dot{u}_{2})=0, \label{eq:1}\\
    &\hat{\cal{D}}_{2}\, u_{2}+f'(\dot{u}_{1}-f{u}_{2}') - \frac{2f^{2}}{r^{2}}(u_{2}-u_{3})  =0, \label{eq:2}\\
    &\hat{\cal{D}}_{2}\, u_{3}+f f' l(l+1)u_{2}=0,  \label{eq:3}\\
    &\hat{\cal{D}}_{2}\, u_{4}=0, \label{eq:4}
\end{align}
where
\begin{equation}
    \hat{\cal{D}}_{2}\equiv \left[-\partial_{t}^{2} + f\partial_{r}(f \partial_{r}) -f\left( \frac{l(l+1)}{r^{2}} + \mu^{2} \right) \right],
\end{equation}
following the convention in \cite{Rosa:2011my}.
The fourth equation \eqref{eq:4} is decoupled to the other three equations, which can be understood from the parity for each sector. One can easily confirm that the decomposition of the vector field \eqref{eq:basis} implies that $\{u_{1},u_{2},u_{3}\}$ are transformed with a factor $(-1)^{l}$ whereas $\{u_{4}\}$ is transformed with $(-1)^{l+1}$ under the parity transformation, $\theta \to \pi - \theta$ and $\phi \to \phi +\pi$. For this reason, we call the former ``even sector'' or ``electric modes'' and the latter ``odd sector'' or ``magnetic modes''. 

Using the Lorentz condition on \eqref{eq:2} again, we can replace it with
\begin{equation}
    \hat{\cal{D}}_{2}\, u_{2}+ \frac{f}{r^{2}} ( r f' - 2f)(u_{2}-u_{3}) =0.  \label{eq:2R}
\end{equation}
Taking $l=0$ for the even sector, corresponding to the {\it monopole mode}, $u_{3}$ is decoupled to the other fields and we obtain a single decoupled equation for the even sector,
\begin{equation}
    \bigg[ -\partial_{t}^{2} + f\partial_{r}(f \partial_{r}) + f \left( \frac{r f' - 2f}{r^{2}} -\mu^{2} \right) \bigg] u_{2} =0. \label{eq:monopole}
\end{equation}
For general $l$, on the other hand, the even sector is described by two coupled equations\footnote{Note that solving \eqref{eq:3} for $u_{2}$ and substituting it into \eqref{eq:2R}, we obtain a single decoupled equation, which is the fourth-order differential equation for $u_{3}$.}, \eqref{eq:2R} and \eqref{eq:3}, for the fields $\{u_{2}, u_{3}\}$. In this paper, we focus on the two decoupled equations, namely the monopole mode in the even sector given by \eqref{eq:monopole} and the odd sector given by \eqref{eq:4}.

Now we assume that the fields are taken to be a plane wave form $u_{i}(t,r) = y_{i}(r) e^{-i \omega t}$ with the frequency $\omega$. Introducing the tortoise coordinate $\dd r_{*} = f^{-1}\dd r$, the equations in question are given by
\begin{align}
    &\bigg[ \partial_{r_{*}}^{2} + \omega^{2}-f \left(  
 \frac{2(r-3)}{r^{3}}+\mu^{2} \right) \bigg] y_{2} =0, \label{eq:EOMy2}\\
 &\bigg[ \partial_{r_{*}}^{2} + \omega^{2}-f \left(  
 \frac{l(l+1)}{r^{2}}+\mu^{2} \right) \bigg] y_{4} =0, \label{eq:EOMy4}
\end{align}
corresponding to the monopole mode in the even sector and the modes in the odd sector, respectively. For our purpose, we further rewrite the equations by introducing
\begin{equation}
    z= \frac{r}{2M}, \quad \Phi_{i} = \sqrt{\frac{z-1}{z}} y_{i},
\end{equation}
then we obtain
\begin{align}
    &\Phi_{2}''(z)+ Q_{2}(z) \Phi_{2}(z) =0, \label{eq:Phi2} \\
    &\Phi_{4}''(z)+ Q_{4}(z) \Phi_{4}(z) =0, \label{eq:Phi4} 
\end{align}
where
\begin{align}
    &Q_{2}(z) = \frac{1}{z^{2}(z-1)^{2}}\sum_{i=0}^{4} A_{2,i}z^{i}, \\
    &Q_{4}(z) = \frac{1}{z^{2}(z-1)^{2}}\sum_{i=0}^{4} A_{4,i}z^{i},
\end{align}
with
\begin{equation}
\begin{aligned}[c]
A_{2,0}&=-\frac{3(4+M)}{4M},\\
A_{2,1}&=3\left(1+\frac{1}{M}\right),\\
A_{2,2}&=-2,\\
A_{2,3}&=4M^{2}\mu^{2},\\
A_{2,4}&=4M^{2}(\omega^{2}-\mu^{2}),
\end{aligned}
\qquad\qquad\qquad
\begin{aligned}[c]
A_{4,0}&=-\frac{3}{4},\\
A_{4,1}&=1+l(l+1),\\
A_{4,2}&=-l(l+1),\\
A_{4,3}&=4M^{2}\mu^{2},\\
A_{4,4}&=4M^{2}(\omega^{2}-\mu^{2}).
\end{aligned}
\label{eq:coefP}
\end{equation}
Our interest is to compute the eigenfrequencies for those two modes by imposing the ingoing-wave boundary conditions, namely $y_{i} \sim e^{- i \omega r_{*} }$, at the BH horizon ($r=r_{\rm H}$). As the other boundary condition imposed at infinity ($r=\infty$), both outgoing- and ingoing- wave boundary conditions are possible. The corresponding modes are referred to as the QNM and quasi-bound state, respectively. The details will be discussed later on.

%%%%%%%%%%%%%%%%%%%%%%%%%%
%%%%%% Proca in Sch %%%%%%
%%%%%%%%%%%%%%%%%%%%%%%%%%
\section{Seiberg-Witten curve} \label{sec:SWcurve}
In this section, we consider a geometrical approach to the aforementioned problems based on the correspondence between the quantum Seiberg-Witten curve and the black hole quasinormal modes. To be concrete, we focus on the $SU(2)$ SW theories with $N_{f}=3$ in four spacetime dimensions. We also discuss the connection formula, corresponding to imposing the boundary conditions in the gravity picture.

\subsection{Confluent Heun equation and the dictionary}
The quantum SW curve for $N_{f}=3$ in a convenient form is given by the wave equation \cite{Aminov:2020yma},
\begin{equation}
\begin{aligned}
\hbar^2 \psi''(z)+\left(\frac{1}{z^{ 2}(z-1)^2}\sum_{i=0}^4 \widehat{A}_i z^i \right)\psi(z)=0, \label{eq:SWeq}
\end{aligned}
\end{equation}
which is the normal form of the confluent Heun equation \cite{decarreau1978formes}.
The coefficients are given by
\begin{equation}
\begin{aligned}
\widehat{A}_0&=-\frac{(m_1-m_2)^2}{4}+\frac{\hbar^2}{4},\\
\widehat{A}_1&=-E-m_1 m_2-\frac{m_3 \Lambda_3}{8}-\frac{\hbar^2}{4},\\
\widehat{A}_2&=E+\frac{3m_3 \Lambda_3}{8}-\frac{\Lambda_3^2}{64}+\frac{\hbar^2}{4},\\
\widehat{A}_3&=-\frac{m_3 \Lambda_3}{4}+\frac{\Lambda_3^2}{32},\\
\widehat{A}_4&=-\frac{\Lambda_3^2}{64} ,
\end{aligned}
\label{eq:coefSW}
\end{equation}
where the gauge parameters $m_1$, $m_2$ and $m_3$ are the (anti-) fundamental hypermultiplets masses, and $\Lambda_3$ denotes the rescaled gauge coupling constant. $E$ is the eigenenergy, and $\hbar$ represents the Planck constant. For later convenience, we set $\hbar=1$.

Since we now obtain the same form of equation as in \eqref{eq:Phi2}-\eqref{eq:Phi4}, the dictionary of the parameters between two different theories can be read off by comparing \eqref{eq:coefP} and \eqref{eq:coefSW}. For the monopole mode in the even sector, we find
\begin{equation}
\begin{aligned}
    & \Lambda_{3} = -16 i M \sqrt{\omega^{2}-\mu^{2}}, \quad E=-\frac{9}{4}+2M^{2}(4\omega^{2}-\mu^{2}), \\
    & m_{1} = \sqrt{1+\frac{3}{M}-4M^{2}\omega^{2} - 4i \omega \sqrt{M(M+3)}}, \\
    & m_{2} =- \frac{1+\frac{3}{M} +4M^{2}\omega^{2}}{\sqrt{1+\frac{3}{M}-4M^{2}\omega^{2} - 4i \omega \sqrt{M(M+3)}}}, \\
    & m_{3} = -\frac{i M(2\omega^{2}-\mu^{2})}{\sqrt{\omega^{2}-\mu^{2}}},
\end{aligned}
\label{eq:dictionary1}
\end{equation}
and for the odd sector we find
\begin{equation}
\begin{aligned}
    & \Lambda_{3} = -16 i M \sqrt{\omega^{2}-\mu^{2}}, \quad E=-\frac{1}{4}-l(l+1)+2M^{2}(4\omega^{2}-\mu^{2}), \\
    & m_{1} = -1-2iM\omega, \\
    & m_{2} = 1-2iM\omega, \\
    & m_{3} = -\frac{i M(2\omega^{2}-\mu^{2})}{\sqrt{\omega^{2}-\mu^{2}}}.
\end{aligned}
\label{eq:dictionary2}
\end{equation}

\subsection{Connection formula}
Now we consider the boundary conditions at the horizon and spatial infinity.
In general, we obtain the following asymptotic behaviors of the field at the horizon and spatial infinity:
\begin{align}
    &y_{i}(r_{*}\to -\infty) \sim e^{- i \omega r_{*}}, \label{eq:inout}\\
    &y_{i}(r_{*}\to +\infty) \sim B_{i}(\omega)e^{-k r_{*}}+C_{i}(\omega)e^{+k r_{*}}, \label{eq:infinity}
\end{align}
where $B_{i}(\omega), C_{i}(\omega)$ are complex coefficients and we define
\begin{equation}
    k = \sqrt{\mu^{2}-\omega^{2}}.
\end{equation}
such that ${\rm Re}[k]>0$.
The sign of the exponent in (\ref{eq:inout}) corresponds to an ingoing wave near the horizon.
At spatial infinity, there are two modes as in \eqref{eq:infinity}, corresponding to a decaying or divergent behavior, respectively.
QNM correspond to ingoing at the horizon and purely outgoing $B_{i}(\omega)=0$ at spatial infinity, whereas the quasi-bound state solutions are ingoing at the horizon and decaying $C_{i}(\omega)=0$ at spatial infinity. The quasi-bound states are spatially localized within the vicinity of the BH because they are exponentially decaying toward spatial infinity.

For the field $\psi(z)$ solving the wave equation \eqref{eq:SWeq} in the SW theory, the asymptotic forms are given by
\begin{align}
    &\psi(z\to 1) \sim (z-1)^{\frac{1}{2} + \frac{m_{1}+m_{2}}{2}}, \\
    &\psi(z\to\infty) \sim \left(\Lambda_{3} z \right)^{\pm m_{3}} e^{\pm \frac{\Lambda_{3}z}{8} }. \label{eq:infinityY}
\end{align}
We choose ingoing wave boundary condition for the asymptotic behavior at the horizon.

Applying the connection formula discussed in \cite{Bonelli:2021uvf, Bonelli:2022ten}, different boundaries are linked by crossing symmetry, and the asymptotic behavior at spatial infinity is given as
\begin{equation}
    \Phi_{i}(z\to\infty) \sim \tilde{B}_{i}(\Lambda_{3}, a,{\bf m}) \left(\Lambda_{3} z \right)^{+ m_{3}} e^{+ \frac{\Lambda_{3}z}{8} } + \tilde{C}_{i}(\Lambda_{3}, a,{\bf m}) \left(\Lambda_{3} z \right)^{-m_{3}} e^{-\frac{\Lambda_{3}z}{8} }, \label{eq:asympt}
\end{equation}
as a linear combination of the terms in (\ref{eq:infinityY}).
We give the definition of $\tilde{B}_{i}$ and $\tilde{C}_{i}$ in Appendix \ref{app:A}.
The QNM or quasi-bound state can then be chosen by setting $\tilde{B}_{i}=0$ or $\tilde{C}_{i}=0$, respectively.
The quantization condition for the QNM frequencies is obtained,
\ba
&& \Pi_{B}^{(3)}\left(E ,{\bf m}, \Lambda_{3} ,\hbar \right)=\partial_a { \mathcal{F}}^{(3)} ({a}, {\bf m},\Lambda_{3}, \hbar)\Big |_{a=a(E,{\bf m},\Lambda_{3}, \hbar)} = 2\pi \left(n+{1\over 2} \right), \label{eq:BforQNM}
\ea
while applying the Matone relation \eqref{matone}.

%%%%%%%%%%%%%%%%%%%%%%%%%%
%%%%%%   Results    %%%%%%
%%%%%%%%%%%%%%%%%%%%%%%%%%
\section{Duality tests} \label{sec:results}
{In the previous sections, we discussed the dictionary of both equations to be solved and boundary conditions, between the gravity theory and SW theory.}
In this section, we turn to compute the frequencies for QNM and quasi-bound states by using the SW curve method according to the dictionary we found. In the following, we show the results for the monopole mode in the even sector and the mode in the odd sector{, and confirm the consistency of the SW curve approach}.
In actual calculations, we consider the instanton counting series in the NS free energy up to the 5th order (7th order in Fig.~\ref{fig:Nvsw} for convengence discussion).

\subsection{Quasinormal modes}
\begin{figure}[tbp]
    \centering
    \includegraphics[width=0.6\linewidth]{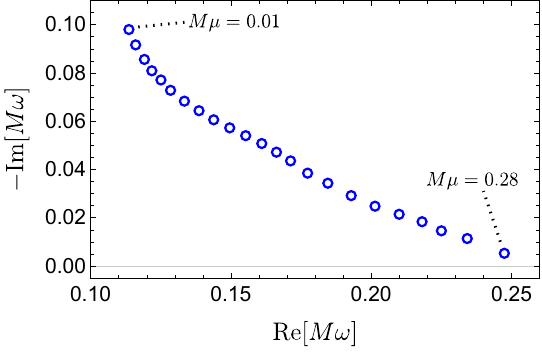}
    \caption{The QNM with $n=0$ for the monopole mode in the even sector. The values of $M\mu$ are varied from $M\mu=0.01$ to $M\mu =0.28$.}
    \label{fig:MonoN0}
\end{figure}

Figure \ref{fig:MonoN0} shows the QNM with $n=0$ for the monopole mode in the even sector, computed by the SW curve. As shown, the imaginary part of the frequency in the lowest QNM approaches to zero as $\mu$ increases, while the real part becomes larger. This indicates the existence of infinitely long living oscillating modes{, which is characteristic for massive vector field perturbations in the asymptotically flat spacetime \cite{Konoplya:2005hr}}. Our result is consistent with the previous numerical calculations (see Figure 4 in \cite{Konoplya:2005hr}).

To check the consistency, we also show several values of the frequencies for the lower QNM ($n=0,1$) in the numerical calculations and the SW curve method in table \ref{tab:MonoTable}. The numerical results are computed by using the continued fraction method \cite{Konoplya:2005hr}, discussed in detail in Appendix \ref{app:C}. We choose $M\mu=0.01$, $M\mu=0.10$, and $M\mu=0.25$ as the vector field mass for each $n$.
{The real part of the second mode frequency ($n=1$) decreases as the mass increases, whereas that of the lowest mode ($n=0$) increases. This implies that the lowest and second mode approach the purely real oscillating mode and the purely imaginary overdamped mode, respectively.}
We consider that the deviation between the two methods stems from the numerical errors due to the truncation in the instanton counting series in the Nekrasov-Shatashvili free energy \cite{Nekrasov:2009rc}. Still, the frequencies with the SW curve method are sufficiently consistent with the numerical results and the same dependence on $M\mu$ is confirmed for both $n=0,1$.
\begin{table}[ht]
    \centering
    \caption{The lower frequencies of the QNM in the numerical calculations and SW curve method. We choose $M\mu=0.01$, $M\mu=0.10$, and $M\mu=0.25$ as the vector field mass.}\vspace{0.5em}
    \includegraphics[width=0.7\linewidth]{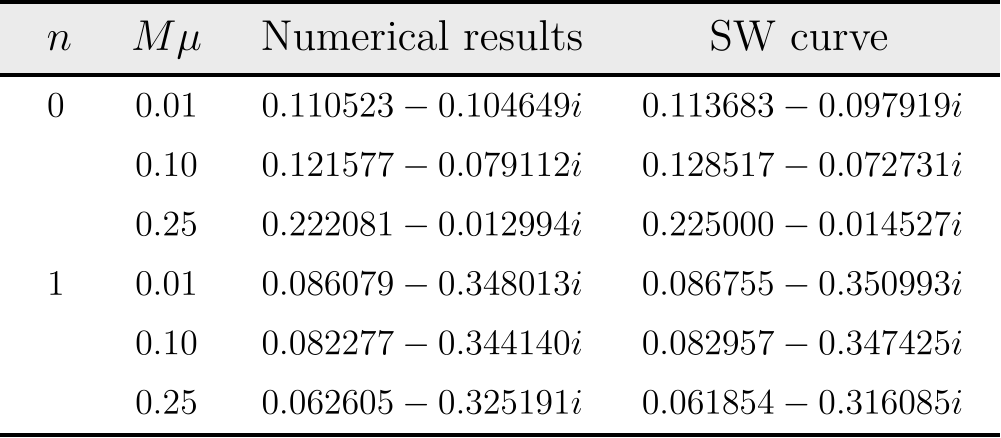}
    \label{tab:MonoTable}
\end{table}

Now we study the QNM for the odd sector. Table~\ref{tab:l12Table} shows the results of $l=1,2$ and $n=0$ obtained by the continued fraction method and SW curve method.
{The modes with non-zero $l$ corresponds to the multipole modes. Both of the dipole ($l=1$) and quadrupole ($l=2$) modes have the same dependence on the mass: the real and imaginary part increase for larger $M{\mu}$. }
Compared to the monopole mode in the even sector, the deviations of the frequencies between the numerical results and SW curve method become larger in the dipole ($l=1$) and quadrupole ($l=2$) modes. However, those should be due to the numerical errors caused by the truncation of the instanton counting series expansion and it is fair to regard that the frequencies with the SW curve method are still consistent with the numerical results.
\begin{table}[ht]
    \centering
    \caption{The lowest frequencies of the QNM in the numerical calculations and SW curve method with $n=0$. We choose $M\mu=0.01$, $M\mu=0.10$, and $M\mu=0.25$ as the vector field mass.}\vspace{0.5em}
    \includegraphics[width=0.7\linewidth]{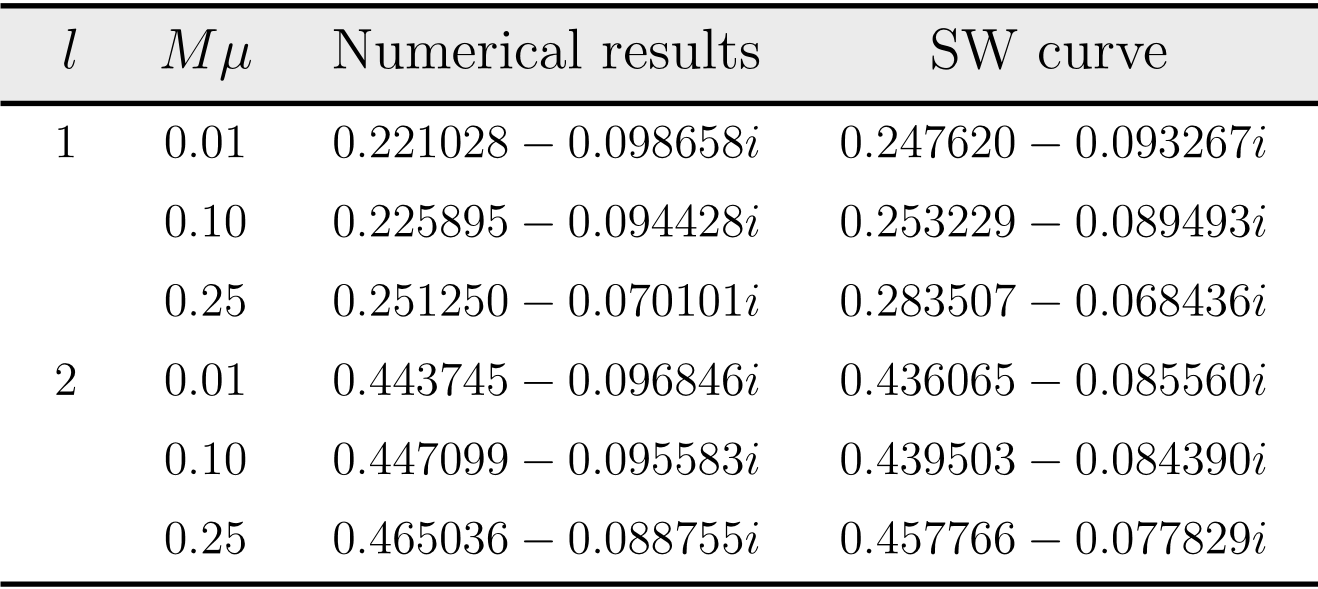}
    \label{tab:l12Table}
\end{table}

\subsection{Quasi-bound states}
Here we confirm that the SW curve method also works for the quasi-bound state. As we studied in \cite{Ge:2024jdx} for the massive scalar field, we have to change the dictionary as $m_{3}\to -m_{3}$ to compute the frequencies for the quasi-bound state. Even in the case of the Proca field, this transformation enables us to compute them from the connection formula.
Table~\ref{tab:QBTable} shows the lowest frequencies of the quasi-bound state for the monopole mode and odd sector mode obtained by the continued fraction method and SW curve method.
As illustrated, we find that the numerical results for the monopole mode and odd sector mode are reproduced by the SW curve method.
{As confirmed in \cite{Rosa:2011my}, the imaginary part of the quasi-bound state frequency monotonically decreases with $M\mu$ and the Proca field slowly decays for smaller mass.}
\begin{table}[ht]
    \centering
    \caption{The lowest frequencies of the quasi-bound state for the monopole mode and odd sector mode in the numerical calculations and SW curve method with $n=0$. We choose $M\mu=0.40$, $M\mu=0.50$, and $M\mu=0.60$ as the vector field mass.}\vspace{0.5em}
    \includegraphics[width=0.75\linewidth]{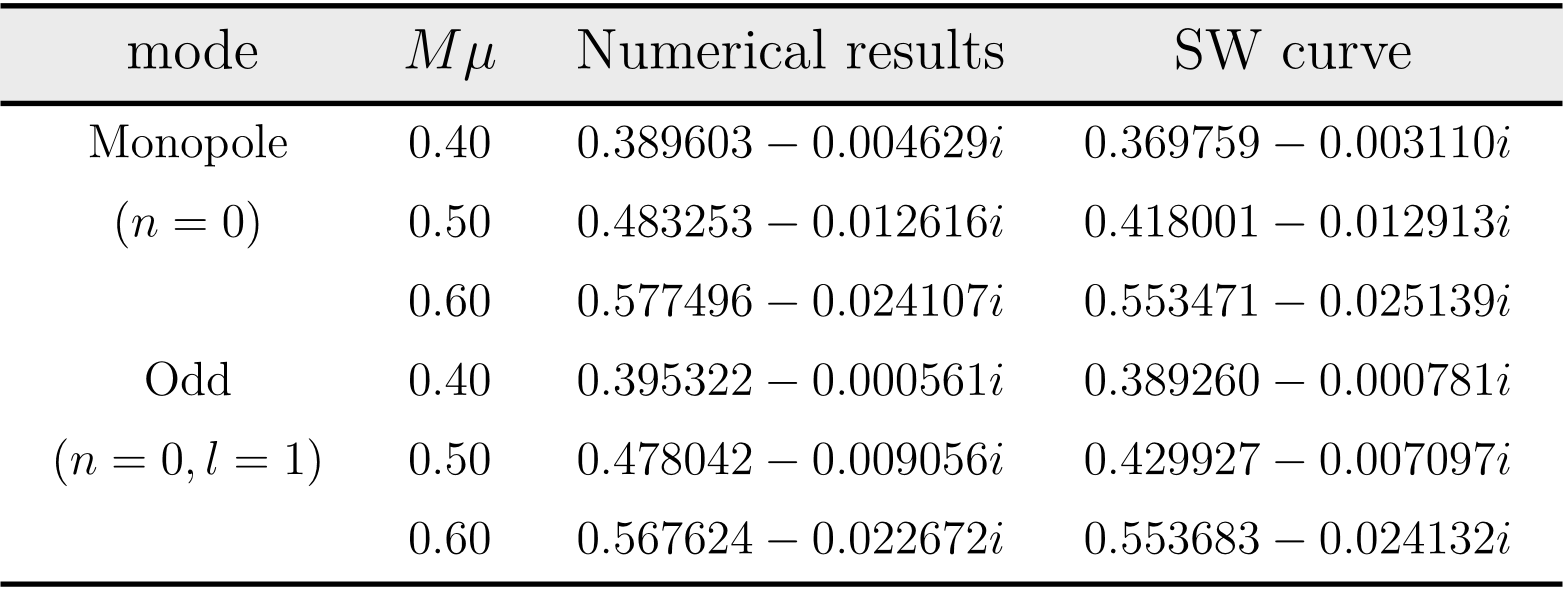}
    \label{tab:QBTable}
\end{table}

\subsection{Convergence of instanton series expansion}
\begin{figure}[tbp]
    \centering
    \includegraphics[width=0.45\linewidth]{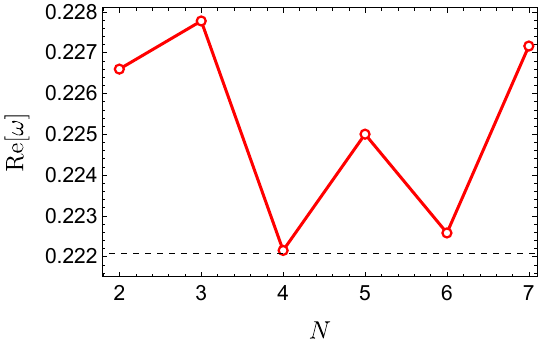}
    \includegraphics[width=0.45\linewidth]{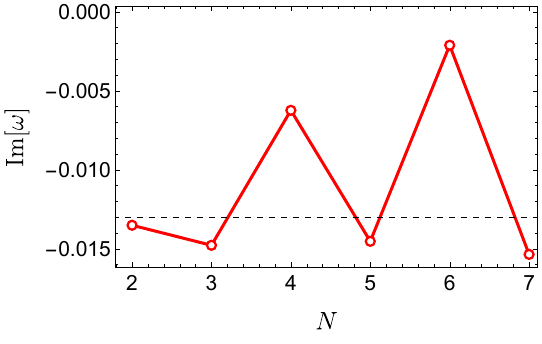}
    \includegraphics[width=0.45\linewidth]{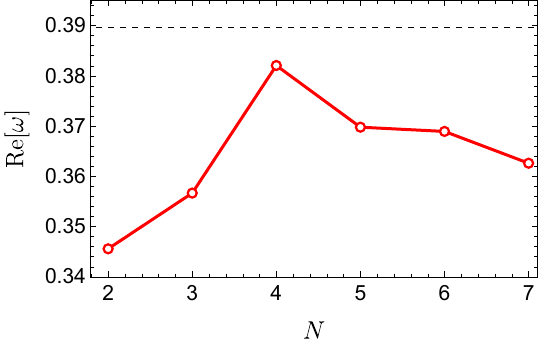}
    \includegraphics[width=0.45\linewidth]{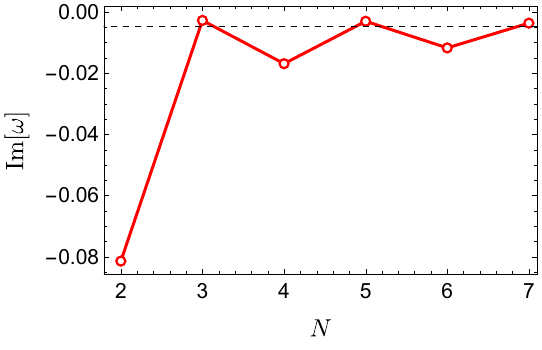}
    \caption{The real (left) and imaginary (right) parts of the lowest-frequency ($n=0$) of the QNMs (upper) and quasi-bound states (lower) for the monopole mode, plotted as functions of the $N$th order in the instanton series expansion. We choose $M\mu = 0.25$ for the QNMs and $M\mu=0.40$ for the quasi-bound states. The black dashed lines denote the numerical results with the continued fraction method.}
    \label{fig:Nvsw}
\end{figure}

 Before concluding this section, we validate the convergence properties of the instanton series expansion. As illustrated in Fig.~\ref{fig:Nvsw}, the real (left) and imaginary (right) components of the lowest-frequency ($n=0$) quasi-normal modes (QNMs, upper panels) and quasi-bound states (lower panels) for the monopole mode are plotted as functions of the $Nth$  order in the instanton series expansion. Our analysis employs $M \mu=0.25$ for QNMs and  $M \mu=0.40$  for quasi-bound states, with numerical benchmarks (black dashed lines) computed via the continued fraction method. While previous results utilized a 5th-order truncation, we extend the expansion to higher orders to probe convergence. Both real and imaginary components exhibit oscillatory behavior around the numerical values as $N$  increases, with a notable exception: the imaginary part of the quasi-bound state frequency demonstrates unambiguous convergence at $N=7$ (which already takes two weeks of computing time on a normal desktop computer), contrasting with the persistent oscillations in other cases.
 This divergence highlights the slow convergence (the convergency is guaranteed in \cite{Its:2014lga, Felder:2017rgg,  Bershtein:2016aef}) inherent to the instanton series expansion, particularly for real frequency components and QNM applications. The observed convergence in the quasi-bound state's imaginary part suggests a hierarchy in the expansion's efficacy across spectral domains, warranting further investigation into resummation techniques or alternative asymptotic expansions for improved accuracy. 

{This result is consistent with
previous studies of SW series convergence; for example,  \cite{DiRusso:2024hmd} show that convergence is typically reached around $N =10$. Since the localization formula used in the paper is known to be inefficient take a rather long time to generate 
  at high order, we leave it for the future work. A promising valuable alternative is to apply a more direct and efficient
method to compute  the free energy introduced in
\cite{Consoli:2022eey}, where this continued fraction
expression allows one to reach higher instanton number much faster.}

On the other hand, the slow convergence of the current method means that we can get a quick and fine estimation with a limited number of instantons. This method may not be the most accurate or efficient way comparing to the known numerical studies,  but it provides fast double checks, and what is more, it implies a new duality, which is meaningful in physics itself.

%%%%%%%%%%%%%%%%%%%%%%%%%%
%%%%%%  Conclusion  %%%%%%
%%%%%%%%%%%%%%%%%%%%%%%%%%

\section{Discussion and conclusion} \label{sec:discussion}
In summary, we investigate the dictionary between the perturbative equations for the massive vector field, i.e. Proca field, in the Schwarzschild spacetime and the quantum SW curve in the supersymmetric gauge field with fundamental multiplets. We apply the original proposition \cite{Aminov:2020yma} into the massive vector fields by extending our previous work in the case of massive scalar fields \cite{Ge:2024jdx}. We focus on the monopole mode in the even sector and the mode in the odd sector, which satisfy the confluent Heun equations, and explore the dictionary to the quantum SW curve. This dictionary enables us to compute the frequencies for the QNM and quasi-bound state consistently within the truncation errors of the instanton counting series. It is fascinating also in the sense of the application to GW physics, since the massive vector boson clouds around BH may introduce unique detectable effects \cite{Baumann:2018vus, Baumann:2019eav} like tidal Love number \cite{Hinderer:2007mb} for future GW observatories \cite{Hu:2017mde,Ruan:2018tsw,TianQin:2015yph,LISA:2017pwj}. 
{The quantum SW method might provide a new perspective for a deeper understanding of such various GW phenomena, instead of the conventional approaches. }

{Before concluding, we briefly remark possible future perspectives.}
Although here in the even sector we focus only on the monopole mode, it would be interesting to extend our analysis into the multipole modes ($l\geq1$) in the even sector that follow the two coupled differential equations \cite{Rosa:2011my}. The existence of the corresponding quantum SW curves is highly non-trivial.
An extension to the Proca field in the Kerr spacetime is also interesting. The perturbative equations are no longer the Heun type, but contain more singularities \cite{Pani:2012bp,Frolov:2018ezx,Baumann:2019eav}. As a result, the corresponding connection formula should be given by the quiver gauge theories, as recently discussed in \cite{Liu:2024eut}. We will present the work for the Kerr case in the near future.

%%%%%%%%%%%%%%%%%%%%%%%%%%
%%%%  Acknowledgment  %%%%
%%%%%%%%%%%%%%%%%%%%%%%%%%
\section*{Acknowledgments}
The authors acknowledge Yutaka Matsuo, Yang Lei, Hongfei Shu and Rui-Dong Zhu for helpful discussions. The work of XHG is supported in part by NSFC, China (Grant No. 12275166 and No. 12311540141). MM is supported by Shanghai Post-doctoral Excellence Program (No.\,2023338). KZ (Hong Zhang) is supported by a classified fund of Shanghai city.

%%%%%%%%%%%%%%%%%%%%%%%%%%
%%%%%%%  Appendix  %%%%%%%
%%%%%%%%%%%%%%%%%%%%%%%%%%
\appendix

\section{Coefficients in the connection formula} \label{app:A}
Here we explicitly show the coefficients in the asymptotic behavior at spatial infinity \eqref{eq:asympt} and discuss the quantization condition to obtain the QNM and quasi-bound state frequencies. More details on the discussion in this appendix was originally presented in our previous work \cite{Ge:2024jdx}.
Following the convention in \cite{Bonelli:2021uvf}, those coefficients are written as
\begin{equation}
        \Phi_{i}(z\to\infty) \sim \tilde{B}_{i}(\Lambda, a,{\bf m}) \left(\Lambda z \right)^{+ m_{3}} e^{+ \frac{\Lambda z}{2} } +\tilde{C}_{i}(\Lambda, a,{\bf m}) \left(\Lambda z \right)^{-m_{3}} e^{-\frac{\Lambda z}{2} },
\end{equation}
with $\Lambda= \Lambda_{3}/4$.
Besides,  $a$ should be replaced by $-i a$ to match our convention, which leads to a new U(1) factor along with a sign difference when defining the free energy $\cF$. Since those coefficients are independent on the index of the field, we omit $i$, and simply write $\tilde{B}$ and $\tilde{C}$ in the following.
They are explicitly given by
\begin{eqnarray}
    &&\begin{split}
    \tilde{B}(\Lambda, a,{\bf m}) = \Lambda^{a} &M_{\alpha_{2+},\alpha_{+}} {\cal{A}}_{\alpha_{+}m_{0+}} 
    \frac{
    \mel{\Delta_{\alpha+},\Lambda_{0},m_{0+}}{V_{\alpha_{2}}(1)}{\Delta_{\alpha_{1}}}
    }
    {
    \mel{\Delta_{\alpha},\Lambda_{0},m_{0}}{V_{\alpha_{2+}}(1)}{\Delta_{\alpha_{1}}}
    } \\
    &+
    \Lambda^{-a} M_{\alpha_{2+},\alpha_{-}} {\cal{A}}_{\alpha_{-}m_{0+}} 
    \frac{
    \mel{\Delta_{\alpha-},\Lambda_{0},m_{0+}}{V_{\alpha_{2}}(1)}{\Delta_{\alpha_{1}}}
    }
    {
    \mel{\Delta_{\alpha},\Lambda_{0},m_{0}}{V_{\alpha_{2+}}(1)}{\Delta_{\alpha_{1}}}
    },
    \end{split} \\
    &&\begin{split}
    \tilde{C}(\Lambda, a,{\bf m}) = \Lambda^{a} &M_{\alpha_{2+},\alpha_{+}} {\cal{A}}_{\alpha_{+}m_{0-}} 
    \frac{
    \mel{\Delta_{\alpha+},\Lambda_{0},m_{0-}}{V_{\alpha_{2}}(1)}{\Delta_{\alpha_{1}}}
    }
    {
    \mel{\Delta_{\alpha},\Lambda_{0},m_{0}}{V_{\alpha_{2+}}(1)}{\Delta_{\alpha_{1}}}
    } \\
    &+
    \Lambda^{-a} M_{\alpha_{2+},\alpha_{-}} {\cal{A}}_{\alpha_{-}m_{0-}} 
    \frac{
    \mel{\Delta_{\alpha-},\Lambda_{0},m_{0-}}{V_{\alpha_{2}}(1)}{\Delta_{\alpha_{1}}}
    }
    {
    \mel{\Delta_{\alpha},\Lambda_{0},m_{0}}{V_{\alpha_{2+}}(1)}{\Delta_{\alpha_{1}}}
    }.
    \end{split}
\end{eqnarray}
Then the subsequent application of the connection formula are totally parallel to the discussion in Appendix B of \cite{Ge:2024jdx}.

\section{Parametrically controlled expansion formulas for the QNM frequencies}  \label{app:B}
As illustrated in \cite{Aminov:2020yma},  we start from 
the U(2) Nekrasov partition function with 3 flavors,
\begin{equation}
Z^{(3)}({a};{\bf m}; \Lambda_{3},\epsilon_1, \epsilon_2)=\sum_{\boldsymbol{Y}} \Biggl({\Lambda_{ 3}\over 4}\Biggr)^{\ell (\boldsymbol{Y})}  \CZ_{\boldsymbol{Y}}^{\rm vect} \CZ_{\boldsymbol{Y}}^{\rm fund}, 
\end{equation}
where  $\boldsymbol{Y}=(Y_1, Y_2)$ denotes a vector Young Tableau,  with $\ell (\boldsymbol{Y})$ its total length,  and ${\bf m}=\{m_1, \cdots, m_{3}\}$.

For a Young Tableau $Y=(y_1, y_2, \cdots)$ and its transpose $ Y^t=(y_1^t, y_2^t, \cdots)$,  the arm-length and and leg-length for a cell $ \square=(i,j) $ are defined respectively by 
\begin{equation}
  A_Y( \square)= y^t_j-i \,,\qquad L_Y( \square)=y_i-j\,,
\end{equation}
then
\ba
\CZ_{\boldsymbol{Y}}^{\rm vect} &&=\prod_{I,J=1}^2 \prod_{\square \in Y_I} {1\over \alpha_I -\alpha_J -\epsilon_1 A_{Y_J}(\square) +\epsilon_2 \left( L_{Y_I}(\square)+1 \right)} \\
&& \times  \prod_{\square\in Y_J}
{1\over \alpha_I -\alpha_J +\epsilon_1 \left(A_{Y_I}(\square)+1\right){-} \epsilon_2 L_{Y_J}(\square)}, \nn
\ea
and
\ba
\CZ_{\boldsymbol{Y}}^{\rm fund} =\prod_{k=1}^{3}\prod_{I=1}^2 \prod_{(i,j) \in Y_I} \left(\alpha_I +m_k+(i-\frac{1}{2})\epsilon_1+(j-\frac{1}{2})\epsilon_2\right).
\ea

The U(2)
 instanton part of Nekrasov-Shatashvili free energy $F_{\rm inst}^{(3)}({a}; {\bf m} ;  \Lambda_{3},\hbar)$ is defined by 
\begin{equation}
F_{\rm inst}^{(3)}({a}; {\bf m} ;  \Lambda_{3},\hbar) = -\hbar \, \lim_{\epsilon_2
  \rightarrow 0} \, \epsilon_2 \log Z^{(3)}({ \ri}{a},   {\bf m} ,  \hbar , \epsilon_2).
\end{equation}

% The Nekrasov partition function $Z^{(N_f)}({ \ri}{a},   {\bf m} ,  \epsilon_1 , \epsilon_2)$ is exact in $\epsilon_i$, and can be written explicitly in terms of $\Lambda_{N_f}$ instanton expansion \cite{Alday:2009aq}, as a convergent series. 

It can be obtained through $\Lambda_{3}$ instanton expansion,
\begin{equation}
\begin{aligned}
F_{\rm inst}^{(3)}({a};  {\bf m};\Lambda_{3} , \hbar) = &\frac{1}{8} \left(-\frac{4 m_1 m_2 m_3}{4 a^2+\hbar ^2}+m_1+m_2+m_3+{  \hbar} \right)\Lambda_3 
+\mathcal{O}(\Lambda_3^2) .
\end{aligned}
\end{equation}
The SU(2) instanton part ${ \mathcal{F}}_{\rm inst}^{(3)}({a};    {\bf m}  ; \Lambda_{3}, \hbar)$ can be obtained by removing the U(1) contribution from the U(2)
 instanton part $F_{\rm inst}^{(3)}({a}; {\bf m} ;  \Lambda_{3},\hbar)$, 
\begin{equation}
{\mathcal{F}}_{\rm inst}^{(3)}({a};  {\bf m};  \Lambda_{3}, \hbar) = F_{\rm inst}^{(3)}({a};  {\bf m};\Lambda_{3} , \hbar) -\frac{\Lambda_3}{8} \left(+m_1+m_2+m_3+{  \hbar} \right).
\end{equation}

The full SU(2) Nekrasov-Shatashvili free energy $ { \mathcal{F}}^{(3)} ({a}; {\bf m};\Lambda_{3}, \hbar)$ \cite{Nekrasov:2009rc} has contributions from its classical, one-loop and instanton components, with 
 this relation given explicitly as \cite{Aminov:2020yma}
\begin{equation}
\begin{aligned}
\label{partialF}
&\partial_a { \mathcal{F}}^{({3})} ({a}; {\bf m};\Lambda_{3}, \hbar) 
= -2\,a\log{\left [\frac{\Lambda_{3} }{4\hbar }\right ]}-\pi\hbar\\
&-2\,\ri\,\hbar\log{\left [
\dfrac{\Gamma({1+\frac{2\ri a}{\hbar})} } {\Gamma(1-\frac{2\ri a}{\hbar})}\right ]}
-\ri\,\hbar
\sum_{j=1}^{3}\log {\left [\dfrac{\Gamma ({\frac12+\frac{m_j-\ri a}{\hbar}})}
{\Gamma ({\frac12+\frac{m_j+\ri a}{\hbar}})}\right ]} +\dfrac{\partial { \mathcal{F}}_{\rm inst}^{(3)}({a};    {\bf m}  ; \Lambda_{3}, \hbar) }{\partial a}. 
\end{aligned}
\end{equation}

To consider the SW spectrum, we can choose the contour as the so-called A-circle and B-circle on a torus, with the quantization condition
\be
\Pi_{I}^{(N_{f})}(E, {\bf m}, \Lambda_{N_f}, \hbar) =N_I  \left( n+\frac{1}{2} \right), \quad I =A,B, \quad n=0,1,2,\cdots,
\ee
where  $N_A =i$ and $N_B =2\pi$.
For the quantum A-period, we have
\be
\label{atoF} 
\Pi_{A}^{(N_f)}(E,{\bf m}, \Lambda_{N_f},\hbar)=a(E,{\bf m},\Lambda_{N_f}, \hbar),
\ee
while the quantum B-period is
\be 
\label{BtoF} 
\Pi_{B}^{(N_f)}(E,{\bf m}, \Lambda_{N_f},\hbar)=\partial_a { \mathcal{F}}^{(N_f)} ({a}, {\bf m},\Lambda_{N_f}, \hbar)\Big |_{a=a(E,{\bf m},\Lambda_{N_f}, \hbar)}.
\ee
Here the derivative of the full Nekrasov-Shatashvili free energy $\partial_a { \mathcal{F}}^{(N_f)} ({a}, {\bf m},\Lambda_{N_f}, \hbar)$ can be obtained through \eqref{partialF}. Notice that $E$ should be inverted as a function of the quantum mirror map
$a$ by the Matone relation \cite{Matone:1995rx}:
\be
\label{matone}
{E=a^2-\dfrac{\Lambda_{N_f}}{4-N_f} \dfrac{\partial \mathcal{F}_{\rm inst}^{(N_f)}(a;{\bf m};\Lambda_{N_f}, \hbar)}{\partial \Lambda_{N_f}}}.
\ee 
In practice, this is also done as an expansion of $\Lambda_{3}$.
We can then impose the quantum B-period to obtain the frequency for the QNM and quasi-bound state, discussed in the main text.

%%%%%%%%%%%%%%
In details, for example,  for the odd sector case with the dictionary  \eqref{eq:dictionary2},
the frequency $\omega$ can be obtained by solving 
\begin{equation}
\label{EaF} 
\partial_a { \mathcal{F}}^{(3)} \left(a,{\bf m}, -16 i M \sqrt{\omega^{2}-\mu^{2}},1\right)\Big |_{a=a(E,{\bf m},\Lambda_{N_f}, \hbar)}=2\,\pi\,\brc{n+{1\over 2}},\quad n=0,1,\dots,
\end{equation}
where $a=a\left(-\frac{1}{4}-l(l+1)+2M^{2}(4\omega^{2}-\mu^{2}),{\bf m}, -16 i M \sqrt{\omega^{2}-\mu^{2}},1\right)$ from the Matone relation \eqref{matone},  
${\bf m}=\{-1-2iM\omega, 1-2iM\omega,  -\frac{i M(2\omega^{2}-\mu^{2})}{\sqrt{\omega^{2}-\mu^{2}}}\}$, and $\hbar=1$.

With given $M$ and $\mu$,  the QNM frequency $\omega$ is thus controlled by the convergent expansion parameter $\Lambda_{3}$ and the adjusting parameters $n$ and $l$, through \eqref{partialF} and \eqref{EaF}.

\section{Continued fraction method} \label{app:C}
Following \cite{Rosa:2011my}, we briefly introduce some details of the continued fraction method, or the so-called Leaver's method \cite{Leaver:1985ax}, for computing the QNM or quasi-bound state frequencies of the Proca field. For more general applications and other numerical methods, one can refer to the review \cite{Konoplya:2011qq}. We consider the following ansatz for the vector field \cite{Rosa:2011my}
\begin{equation}
    y_{i}(\omega,r) = f^{-2i\omega}r^{-\nu}e^{q r} \sum_{n} a_{n}^{(i)} [f(r)]^n, \label{eq:CFansatz}
\end{equation}
with $\nu= (\omega^{2}-q^{2})/q^{2}$ and $q=\pm k = \pm\sqrt{\mu^{2}-\omega^{2}}$. The choice of the sign in $q$ corresponds to imposing the boundary condition at infinity; $q=k$ for the QNM and $q=-k$ for the quasi-bound state. Inserting \eqref{eq:CFansatz} into the equations, \eqref{eq:EOMy2} and \eqref{eq:EOMy4}, we obtain that
\begin{align}
    &\alpha_{0} a_{1} + \beta_{0} a_{0} = 0, \\
    &\alpha_{n} a_{n+1} + \beta_{n} a_{n} + \gamma_{n} a_{n-1} =0, \quad (n>0),
\end{align}
where the coefficients $(\alpha_{n}, \beta_{n}, \gamma_{n})$ are given as the functions of $(\omega, q)$ as well as the quantum numbers ($n,l$). The QNM or quasi-bound state frequencies are those for which the following relation holds,
\begin{equation}
    \beta_{n} - \frac{\alpha_{n-1} \gamma_{n}}{\beta_{n-1} - \frac{\alpha_{n-2}\gamma_{n-1}}{\beta_{n-2}- \cdots}} = \frac{\alpha_{n} \gamma_{n+1}}{\beta_{n+1} - \frac{\alpha_{n+1}\gamma_{n+2}}{\beta_{n+2}-\cdots}}.
\end{equation}

For the monopole mode in the even sector, the corresponding coefficients are 
\begin{align}
    & \alpha_{n} = (n+1)(n+1 - 4i\omega), \\
    &\begin{aligned}\beta_{n} = 
    -2n^{2}&-\frac{2}{q}\bigg(\omega^{2}-3q^{2}+(1-4i\omega)q \bigg)n \\
    &+ \frac{1}{q} \bigg( 4i\omega^{3}+(12q-1)\omega^{2} -4iq(3q -1)\omega -4q^{3} + 3q^{2} + q \bigg), 
    \end{aligned}\\
    & \gamma_{n} = n^{2} +\frac{2}{q}\bigg( \omega^{2}-q^{2}+2i\omega q \bigg)n + \frac{1}{q^{2}}(\omega - iq)^{4} -4.
\end{align}
For the modes in the odd sector, the corresponding coefficients are
\begin{align}
    & \alpha_{n} = (n+1)(n+1 - 4i\omega), \\
    &\begin{aligned}\beta_{n} = 
    -2n^{2}&-\frac{2}{q}\bigg(\omega^{2}-3q^{2}+(1-4i\omega)q \bigg)n \\
    &+ \frac{1}{q} \bigg( 4i\omega^{3}+(12q-1)\omega^{2} -4iq(3q -1)\omega -4q^{3} + 3q^{2} - l(l+1)q\bigg), 
    \end{aligned}\\
    & \gamma_{n} = n^{2} +\frac{2}{q}\bigg( \omega^{2}-q^{2}+2i\omega q \bigg)n + \frac{1}{q^{2}}(\omega - iq)^{4} -1.
\end{align}

\bibliography{GMZII}

\providecommand{\href}[2]{#2}\begingroup\raggedright\begin{thebibliography}{10}

\bibitem{GW150914}
{\scshape LIGO Scientific Collaboration and Virgo Collaboration} collaboration,
  B.~P. Abbott, R.~Abbott, T.~D. Abbott, M.~R. Abernathy, F.~Acernese,
  K.~Ackley et~al., \emph{Observation of gravitational waves from a binary
  black hole merger},
  \href{http://dx.doi.org/10.1103/PhysRevLett.116.061102}{\emph{Phys. Rev.
  Lett.} {\bf 116} (Feb, 2016) 061102}.

\bibitem{Newton:1687eqk}
I.~Newton, \emph{{Philosophi\ae{} Naturalis Principia Mathematica}}.
\newblock England, 1687.

\bibitem{vishveshwara1970scattering}
C.~Vishveshwara, \emph{Scattering of gravitational radiation by a schwarzschild
  black-hole}, {\emph{Nature} {\bf 227} (1970) 936--938}.

\bibitem{Kokkotas:1999bd}
K.~D. Kokkotas and B.~G. Schmidt, \emph{{Quasinormal modes of stars and black
  holes}}, \href{http://dx.doi.org/10.12942/lrr-1999-2}{\emph{Living Rev. Rel.}
  {\bf 2} (1999) 2}, [\href{http://arxiv.org/abs/gr-qc/9909058}{{\tt
  gr-qc/9909058}}].

\bibitem{Berti:2009kk}
E.~Berti, V.~Cardoso and A.~O. Starinets, \emph{{Quasinormal modes of black
  holes and black branes}},
  \href{http://dx.doi.org/10.1088/0264-9381/26/16/163001}{\emph{Class. Quant.
  Grav.} {\bf 26} (2009) 163001}, [\href{http://arxiv.org/abs/0905.2975}{{\tt
  0905.2975}}].

\bibitem{Konoplya:2011qq}
R.~A. Konoplya and A.~Zhidenko, \emph{{Quasinormal modes of black holes: From
  astrophysics to string theory}},
  \href{http://dx.doi.org/10.1103/RevModPhys.83.793}{\emph{Rev. Mod. Phys.}
  {\bf 83} (2011) 793--836}, [\href{http://arxiv.org/abs/1102.4014}{{\tt
  1102.4014}}].

\bibitem{Horowitz:1999jd}
G.~T. Horowitz and V.~E. Hubeny, \emph{{Quasinormal modes of AdS black holes
  and the approach to thermal equilibrium}},
  \href{http://dx.doi.org/10.1103/PhysRevD.62.024027}{\emph{Phys. Rev. D} {\bf
  62} (2000) 024027}, [\href{http://arxiv.org/abs/hep-th/9909056}{{\tt
  hep-th/9909056}}].

\bibitem{KalyanaRama:1999zj}
S.~Kalyana~Rama and B.~Sathiapalan, \emph{{On the role of chaos in the AdS /
  CFT connection}},
  \href{http://dx.doi.org/10.1142/S0217732399002777}{\emph{Mod. Phys. Lett. A}
  {\bf 14} (1999) 2635--2648}, [\href{http://arxiv.org/abs/hep-th/9905219}{{\tt
  hep-th/9905219}}].

\bibitem{Brito:2015oca}
R.~Brito, V.~Cardoso and P.~Pani, \emph{{Superradiance}: {New Frontiers in
  Black Hole Physics}},
  \href{http://dx.doi.org/10.1007/978-3-319-19000-6}{\emph{Lect. Notes Phys.}
  {\bf 906} (2015) pp.1--237}, [\href{http://arxiv.org/abs/1501.06570}{{\tt
  1501.06570}}].

\bibitem{East:2017ovw}
W.~E. East and F.~Pretorius, \emph{{Superradiant Instability and Backreaction
  of Massive Vector Fields around Kerr Black Holes}},
  \href{http://dx.doi.org/10.1103/PhysRevLett.119.041101}{\emph{Phys. Rev.
  Lett.} {\bf 119} (2017) 041101}, [\href{http://arxiv.org/abs/1704.04791}{{\tt
  1704.04791}}].

\bibitem{Essig:2013lka}
R.~Essig et~al., \emph{{Working Group Report: New Light Weakly Coupled
  Particles}},  in \emph{{Snowmass 2013}: {Snowmass on the Mississippi}}, 10,
  2013.
\newblock \href{http://arxiv.org/abs/1311.0029}{{\tt 1311.0029}}.

\bibitem{Baumann:2018vus}
D.~Baumann, H.~S. Chia and R.~A. Porto, \emph{{Probing Ultralight Bosons with
  Binary Black Holes}},
  \href{http://dx.doi.org/10.1103/PhysRevD.99.044001}{\emph{Phys. Rev. D} {\bf
  99} (2019) 044001}, [\href{http://arxiv.org/abs/1804.03208}{{\tt
  1804.03208}}].

\bibitem{Detweiler:1980uk}
S.~L. Detweiler, \emph{{KLEIN-GORDON EQUATION AND ROTATING BLACK HOLES}},
  \href{http://dx.doi.org/10.1103/PhysRevD.22.2323}{\emph{Phys. Rev. D} {\bf
  22} (1980) 2323--2326}.

\bibitem{Frolov:2018ezx}
V.~P. Frolov, P.~Krtous, D.~Kubiznak and J.~E. Santos, \emph{{Massive Vector
  Fields in Rotating Black-Hole Spacetimes: Separability and Quasinormal
  Modes}}, \href{http://dx.doi.org/10.1103/PhysRevLett.120.231103}{\emph{Phys.
  Rev. Lett.} {\bf 120} (2018) 231103},
  [\href{http://arxiv.org/abs/1804.00030}{{\tt 1804.00030}}].

\bibitem{Aminov:2020yma}
G.~Aminov, A.~Grassi and Y.~Hatsuda, \emph{{Black Hole Quasinormal Modes and
  Seiberg\textendash{}Witten Theory}},
  \href{http://dx.doi.org/10.1007/s00023-021-01137-x}{\emph{Annales Henri
  Poincare} {\bf 23} (2022) 1951--1977},
  [\href{http://arxiv.org/abs/2006.06111}{{\tt 2006.06111}}].

\bibitem{Seiberg:1994rs}
N.~Seiberg and E.~Witten, \emph{{Electric - magnetic duality, monopole
  condensation, and confinement in N=2 supersymmetric Yang-Mills theory}},
  \href{http://dx.doi.org/10.1016/0550-3213(94)90124-4}{\emph{Nucl. Phys. B}
  {\bf 426} (1994) 19--52}, [\href{http://arxiv.org/abs/hep-th/9407087}{{\tt
  hep-th/9407087}}].

\bibitem{Seiberg:1994aj}
N.~Seiberg and E.~Witten, \emph{{Monopoles, duality and chiral symmetry
  breaking in N=2 supersymmetric QCD}},
  \href{http://dx.doi.org/10.1016/0550-3213(94)90214-3}{\emph{Nucl. Phys. B}
  {\bf 431} (1994) 484--550}, [\href{http://arxiv.org/abs/hep-th/9408099}{{\tt
  hep-th/9408099}}].

\bibitem{Heun:1888}
K.~Heun, \emph{{Zur Theorie der Riemann'schen Functionen zweiter Ordnung mit
  vier Verzweigungspunkten}},
  \href{http://dx.doi.org/10.1007/BF01443849}{\emph{Math. Ann.} {\bf 33} (1888)
  161–179}.

\bibitem{Lei:2023mqx}
Y.~Lei, H.~Shu, K.~Zhang and R.-D. Zhu, \emph{{Quasinormal modes of C-metric
  from SCFTs}}, \href{http://dx.doi.org/10.1007/JHEP02(2024)140}{\emph{JHEP}
  {\bf 02} (2024) 140}, [\href{http://arxiv.org/abs/2308.16677}{{\tt
  2308.16677}}].

\bibitem{Alday:2009aq}
L.~F. Alday, D.~Gaiotto and Y.~Tachikawa, \emph{{Liouville Correlation
  Functions from Four-dimensional Gauge Theories}},
  \href{http://dx.doi.org/10.1007/s11005-010-0369-5}{\emph{Lett. Math. Phys.}
  {\bf 91} (2010) 167--197}, [\href{http://arxiv.org/abs/0906.3219}{{\tt
  0906.3219}}].

\bibitem{Ge:2024jdx}
X.-H. Ge, M.~Matsumoto and K.~Zhang, \emph{{Duality between Seiberg-Witten
  theory and black hole superradiance}},
  \href{http://dx.doi.org/10.1007/JHEP05(2024)336}{\emph{JHEP} {\bf 05} (2024)
  336}, [\href{http://arxiv.org/abs/2402.17441}{{\tt 2402.17441}}].

\bibitem{Witten:1997sc}
E.~Witten, \emph{{Solutions of four-dimensional field theories via M-theory}},
  \href{http://dx.doi.org/10.1201/9781482268737-38}{\emph{Nucl. Phys. B} {\bf
  500} (1997) 3--42}, [\href{http://arxiv.org/abs/hep-th/9703166}{{\tt
  hep-th/9703166}}].

\bibitem{Nekrasov:2012xe}
N.~Nekrasov and V.~Pestun, \emph{{Seiberg-Witten Geometry of Four-Dimensional
  $\mathcal N=2$ Quiver Gauge Theories}},
  \href{http://dx.doi.org/10.3842/SIGMA.2023.047}{\emph{SIGMA} {\bf 19} (2023)
  047}, [\href{http://arxiv.org/abs/1211.2240}{{\tt 1211.2240}}].

\bibitem{Bonelli:2021uvf}
G.~Bonelli, C.~Iossa, D.~P. Lichtig and A.~Tanzini, \emph{{Exact solution of
  Kerr black hole perturbations via CFT2 and instanton counting: Greybody
  factor, quasinormal modes, and Love numbers}},
  \href{http://dx.doi.org/10.1103/PhysRevD.105.044047}{\emph{Phys. Rev. D} {\bf
  105} (2022) 044047}, [\href{http://arxiv.org/abs/2105.04483}{{\tt
  2105.04483}}].

\bibitem{Bonelli:2022ten}
G.~Bonelli, C.~Iossa, D.~Panea~Lichtig and A.~Tanzini, \emph{{Irregular
  Liouville Correlators and Connection Formulae for Heun Functions}},
  \href{http://dx.doi.org/10.1007/s00220-022-04497-5}{\emph{Commun. Math.
  Phys.} {\bf 397} (2023) 635--727},
  [\href{http://arxiv.org/abs/2201.04491}{{\tt 2201.04491}}].

\bibitem{Bianchi:2021mft}
M.~Bianchi, D.~Consoli, A.~Grillo and J.~F. Morales, \emph{{More on the SW-QNM
  correspondence}},
  \href{http://dx.doi.org/10.1007/JHEP01(2022)024}{\emph{JHEP} {\bf 01} (2022)
  024}, [\href{http://arxiv.org/abs/2109.09804}{{\tt 2109.09804}}].

\bibitem{Consoli:2022eey}
D.~Consoli, F.~Fucito, J.~F. Morales and R.~Poghossian, \emph{{CFT description
  of BH{\textquoteright}s and ECO{\textquoteright}s: QNMs, superradiance,
  echoes and tidal responses}},
  \href{http://dx.doi.org/10.1007/JHEP12(2022)115}{\emph{JHEP} {\bf 12} (2022)
  115}, [\href{http://arxiv.org/abs/2206.09437}{{\tt 2206.09437}}].

\bibitem{Cipriani:2025ikx}
A.~Cipriani, G.~Di~Russo, F.~Fucito, J.~F. Morales, H.~Poghosyan and
  R.~Poghossian, \emph{{Resumming Post-Minkowskian and Post-Newtonian
  gravitational waveform expansions}},
  \href{http://arxiv.org/abs/2501.19257}{{\tt 2501.19257}}.

\bibitem{Leaver:1985ax}
E.~W. Leaver, \emph{{An Analytic representation for the quasi normal modes of
  Kerr black holes}},
  \href{http://dx.doi.org/10.1098/rspa.1985.0119}{\emph{Proc. Roy. Soc. Lond.
  A} {\bf 402} (1985) 285--298}.

\bibitem{Rosa:2011my}
J.~G. Rosa and S.~R. Dolan, \emph{{Massive vector fields on the Schwarzschild
  spacetime: quasi-normal modes and bound states}},
  \href{http://dx.doi.org/10.1103/PhysRevD.85.044043}{\emph{Phys. Rev. D} {\bf
  85} (2012) 044043}, [\href{http://arxiv.org/abs/1110.4494}{{\tt 1110.4494}}].

\bibitem{decarreau1978formes}
A.~Decarreau, D.~L. MC, P.~Maroni, A.~Robert and A.~Ronveaux, \emph{{FORMES
  CANONIQUES DES EQUATIONS CONFLUENTES DE L'EQUATION DE HEUN}}, {\emph{Ann.
  Soc. Sci. Bruxelles} {\bf 92} (1978) 53–78}.

\bibitem{Konoplya:2005hr}
R.~A. Konoplya, \emph{{Massive vector field perturbations in the Schwarzschild
  background: Stability and unusual quasinormal spectrum}},
  \href{http://dx.doi.org/10.1103/PhysRevD.73.024009}{\emph{Phys. Rev. D} {\bf
  73} (2006) 024009}, [\href{http://arxiv.org/abs/gr-qc/0509026}{{\tt
  gr-qc/0509026}}].

\bibitem{Nekrasov:2009rc}
N.~A. Nekrasov and S.~L. Shatashvili, \emph{{Quantization of Integrable Systems
  and Four Dimensional Gauge Theories}},  in \emph{{16th International Congress
  on Mathematical Physics}}, pp.~265--289, 8, 2009.
\newblock \href{http://arxiv.org/abs/0908.4052}{{\tt 0908.4052}}.
\newblock \href{http://dx.doi.org/10.1142/9789814304634_0015}{DOI}.

\bibitem{Its:2014lga}
A.~Its, O.~Lisovyy and Y.~Tykhyy, \emph{{Connection Problem for the
  Sine-Gordon/Painlev\'e III Tau Function and Irregular Conformal Blocks}},
  \href{http://dx.doi.org/10.1093/imrn/rnu209}{\emph{Int. Math. Res. Not.} {\bf
  2015} (2015) 8903--8924}, [\href{http://arxiv.org/abs/1403.1235}{{\tt
  1403.1235}}].

\bibitem{Felder:2017rgg}
G.~Felder and M.~M\"uller-Lennert, \emph{{Analyticity of Nekrasov Partition
  Functions}}, \href{http://dx.doi.org/10.1007/s00220-018-3270-1}{\emph{Commun.
  Math. Phys.} {\bf 364} (2018) 683--718},
  [\href{http://arxiv.org/abs/1709.05232}{{\tt 1709.05232}}].

\bibitem{Bershtein:2016aef}
M.~A. Bershtein and A.~I. Shchechkin, \emph{{q-deformed Painlev\'e $\tau$
  function and q-deformed conformal blocks}},
  \href{http://dx.doi.org/10.1088/1751-8121/aa5572}{\emph{J. Phys. A} {\bf 50}
  (2017) 085202}, [\href{http://arxiv.org/abs/1608.02566}{{\tt 1608.02566}}].

\bibitem{DiRusso:2024hmd}
G.~Di~Russo, F.~Fucito and J.~F. Morales, \emph{{Tidal resonances for
  fuzzballs}}, \href{http://dx.doi.org/10.1007/JHEP04(2024)149}{\emph{JHEP}
  {\bf 04} (2024) 149}, [\href{http://arxiv.org/abs/2402.06621}{{\tt
  2402.06621}}].

\bibitem{Baumann:2019eav}
D.~Baumann, H.~S. Chia, J.~Stout and L.~ter Haar, \emph{{The Spectra of
  Gravitational Atoms}},
  \href{http://dx.doi.org/10.1088/1475-7516/2019/12/006}{\emph{JCAP} {\bf 12}
  (2019) 006}, [\href{http://arxiv.org/abs/1908.10370}{{\tt 1908.10370}}].

\bibitem{Hinderer:2007mb}
T.~Hinderer, \emph{{Tidal Love numbers of neutron stars}},
  \href{http://dx.doi.org/10.1086/533487}{\emph{Astrophys. J.} {\bf 677} (2008)
  1216--1220}, [\href{http://arxiv.org/abs/0711.2420}{{\tt 0711.2420}}].

\bibitem{Hu:2017mde}
W.-R. Hu and Y.-L. Wu, \emph{{The Taiji Program in Space for gravitational wave
  physics and the nature of gravity}},
  \href{http://dx.doi.org/10.1093/nsr/nwx116}{\emph{Natl. Sci. Rev.} {\bf 4}
  (2017) 685--686}.

\bibitem{Ruan:2018tsw}
W.-H. Ruan, Z.-K. Guo, R.-G. Cai and Y.-Z. Zhang, \emph{{Taiji program:
  Gravitational-wave sources}},
  \href{http://dx.doi.org/10.1142/S0217751X2050075X}{\emph{Int. J. Mod. Phys.
  A} {\bf 35} (2020) 2050075}, [\href{http://arxiv.org/abs/1807.09495}{{\tt
  1807.09495}}].

\bibitem{TianQin:2015yph}
{\scshape TianQin} collaboration, J.~Luo et~al., \emph{{TianQin: a space-borne
  gravitational wave detector}},
  \href{http://dx.doi.org/10.1088/0264-9381/33/3/035010}{\emph{Class. Quant.
  Grav.} {\bf 33} (2016) 035010}, [\href{http://arxiv.org/abs/1512.02076}{{\tt
  1512.02076}}].

\bibitem{LISA:2017pwj}
{\scshape LISA} collaboration, P.~Amaro-Seoane et~al., \emph{{Laser
  Interferometer Space Antenna}},  \href{http://arxiv.org/abs/1702.00786}{{\tt
  1702.00786}}.

\bibitem{Pani:2012bp}
P.~Pani, V.~Cardoso, L.~Gualtieri, E.~Berti and A.~Ishibashi,
  \emph{{Perturbations of slowly rotating black holes: massive vector fields in
  the Kerr metric}},
  \href{http://dx.doi.org/10.1103/PhysRevD.86.104017}{\emph{Phys. Rev. D} {\bf
  86} (2012) 104017}, [\href{http://arxiv.org/abs/1209.0773}{{\tt 1209.0773}}].

\bibitem{Liu:2024eut}
P.~Liu and R.-D. Zhu, \emph{{Notes on Quasinormal Modes of charged de Sitter
  Blackholes from Quiver Gauge Theories}},
  \href{http://arxiv.org/abs/2412.18359}{{\tt 2412.18359}}.

\bibitem{Matone:1995rx}
M.~Matone, \emph{{Instantons and recursion relations in N=2 SUSY gauge
  theory}}, \href{http://dx.doi.org/10.1016/0370-2693(95)00920-G}{\emph{Phys.
  Lett. B} {\bf 357} (1995) 342--348},
  [\href{http://arxiv.org/abs/hep-th/9506102}{{\tt hep-th/9506102}}].

\end{thebibliography}\endgroup
\bibliographystyle{JHEP}

\end{document}